\documentclass[useAMS,usenatbib]{mn2e} 
\usepackage{epsfig}

\newif\ifAMStwofonts

\newcommand{\target}{A0620--00}

\newcommand{\Msun}{\,$\rm M_{\odot}$}
\newcommand{\Rsun}{\,$\rm R_{\odot}$}

\newcommand{\RL}{\,$R_{\rm L1}$\,}
\newcommand{\kms}{\,$\rm km\,s^{-1}$}
\newcommand{\bh}{\,$M_{\rm 1}$}
\newcommand{\sstar}{\,$M_{\rm 2}$}
\title[Optical spectroscopy of flares from the black hole 
X-ray transient \target\ in quiescence] 
      {Optical spectroscopy of flares from the black hole 
X-ray transient \target\ in quiescence} 

\author[T.\,Shahbaz et al.]
       {T.\,Shahbaz$^{1}$\thanks{E-mail: tsh@ll.iac.es}
        R.I.\,Hynes$^2$,
        P.A.\,Charles$^3$ 
        C.\,Zurita$^4$,
        J.\,Casares$^1$ 
        C.\,A.\,Haswell$^5$,
	\newauthor
        S.\,Araujo-Betancor$^3$,
	C.\,Powell$^5$ \\
$^1$Instituto de Astrof\'\i{}sica de Canarias, 38200 La Laguna,
    Tenerife, Spain \\
$^2$Astronomy Department, The University of Texas at Austin, 1
       University Station C1400, Austin, Texas 78712-0259, USA \\
$^3$Department of Physics and Astronomy, University of Southampton, 
    Southampton, SO17 1BJ, UK \\
$^4$Centro de Astronomia e Astrof\'\i{}sica da Universidade de Lisboa,
       Tapada da Ajuda, 1314-018, Lisboa, Portigal \\
$^5$Department of Physics and Astronomy, The Open University, Walton
    Hall, Milton Keynes, MK7 6AA}

\pagerange{\pageref{firstpage}--\pageref{lastpage}}
\pubyear{2003}

\begin{document}
\maketitle
\begin{abstract}
\noindent

We present a time-resolved spectrophotometric study of the optical variability
in the quiescent soft X-ray transient \target. Superimposed on the
double-humped continuum lightcurve are the well known flare events which last
tens of minutes.  Some of the flare events that appear in the continuum
lightcurve are also present in the emission line lightcurves.  From the Balmer
line flux and variations, we find that the persistent emission is optically
thin. During the flare event at phase 1.15 the Balmer decrement dropped
suggesting either a significant increase in temperature or that the flares are 
more optically thick than the continuum.
The data suggests that there are two HI emitting regions, the accretion disc 
and the accretion stream/disc region, with different Balmer decrements. 
The orbital modulation of H$\alpha$ with the continuum suggests that the
steeper decrement is most likely associated with the stream/disc impact region.

By isolating the flare's spectrum we find that it has a frequency power-law
index of --1.40$\pm$0.20 (90 percent confidence). The flare spectrum can also 
be described by an optically thin gas with a temperature in the range 
10000--14000\,K that covers 0.05--0.08 percent (90 percent confidence) of the 
accretion disc's surface. Given these parameters, the possibility that the 
flares arise from the bright-spot cannot be ruled out.

We construct Doppler images of the H$\alpha$ and H$\beta$ emission lines. Apart
from showing enhanced blurred emission at the region where the stream impacts 
the  accretion disc, the maps also show significant extended structure from  the
opposite side of the disc. The trailed spectra show characteristic S-wave 
features that can be interpreted in the context of an eccentric accretion disc.

\end{abstract}

\section{Introduction}
\label{INTRODUCTION}

Soft  X-ray transients  (SXTs)  are  a subset of  low-mass X-ray  binaries
(LMXB) that  display episodic, dramatic X-ray  and optical  outbursts, which usually
last for several months.  In the interim the  SXTs  are  in  a state  of
quiescence during which the optical  emission is dominated  by the luminosity 
of the faint  companion star \citep{Paradijs95}. In quiescence the  optical
lightcurves  exhibit the classical double-humped ellipsoidal modulation, which
is  due to the differing  aspects that the  tidally distorted secondary star
presents to the observer throughout its orbit (e.g. see \citealt{Avni75}; 
\citealt{Wilson71}; \citealt{Tjemkes86} and \citealt{Shahbaz03a}).

The LMXB SXT prototype, \target\ (=V616\,Mon) was discovered in 1975 by
the Ariel-5 satellite \citep{Elvis75} during an X-ray outburst whose peak
flux made it the brightest nonsolar source ever seen. The optical  counterpart 
was subsequently identified \citep{Boley75} and after the system had faded back 
to its quiescent level,  \target\ was found to be a 7.8\,hr binary system  
containing a K-type secondary star and a non-stellar continuum source attributed 
to an accretion disc  (\citealt{Oke77}; \citealt{McClintock83}). A radial velocity 
study led to the discovery of the first black hole primary in an X-ray transient 
\citep{MR86}. Thereafter, many UV/optical/infrared studies of \target\ in quiescence  
have been undertaken (\citealt{Haswell93}; \citealt{Marsh94}; 
Shahbaz, Naylor \& Charles 1994; McClintock, Horne \& Remillard 1995; 
\citealt{Shahbaz99}; \citealt{Froning01}; Gelino, Harrison \& Orosz 2001), 
that have enabled the system parameters to be
determined. The K3--K4V  secondary star has a  radial velocity semi-amplitude of
433$\pm$3\kms which gives a mass function of 2.72$\pm$0.06\Msun
\citep{Marsh94}, the inclination is $i$=41$\pm$3$^\circ$ (\citealt{Shahbaz94};
\citealt{Gelino01}), the binary  mass ratio $q$ is(\sstar/\bh)=0.067
(\citealt{Marsh94}; where \bh\ and \sstar\ are the  masses of the compact
object and secondary star respectively) and  \bh=11$\pm$1.9\Msun
(\citealt{Shahbaz94}; \citealt{Gelino01}).

The high quiescent X-ray luminosity of \target\ compared to other black hole 
binaries indicates that it is not completely dormant \citep{MHR95}, but that
there  is still some continuing accretion, although at a very low rate.
Superposed on the ellipsoidal modulation of the secondary star  are many rapid
flares on  timescales of tens of minutes or less (\citealt{Haswell92};
\citealt{Zurita03};  \citealt{Hynes03}), which appear to be a common feature 
in quiescent black hole and neutron star   X-ray transients \citep{Zurita03}. 
The results obtained from timing analyses alone have not proven conclusive and
the origin of the variability  still remains uncertain. The most likely
explanations  are magnetic reconnection  events in the disc, optical  emission
from an  advective region, reprocessed X-ray variability and flickering from
the accretion stream impact point \citep{Zurita03}.  A spectrophotometric study
of the flares in V404\,Cyg showed that the continuum and H$\alpha$ emission
line were at times correlated \citep{Hynes02}.  The kinematics of the 
H$\alpha$ flare events suggest that the whole line profile  participates in the
flare, and can be explained  if the optical flare is induced by variable
photoionisation from the X-ray source. 

Here we report on our time-resolved optical spectrophotometric study of 
\target\ in quiescence. We study the Balmer line and continuum short-term 
variations and examine the spectrum of the variability. We infer the physical
parameters and origin of the short-term flare. Finally  the low resolution
spectra provides us with the kinematic resolution  to study the emission  line
variations using the method of Doppler tomography.

\begin{table}
\caption{Log of observations.}
\label{Table:log}
\begin{center}
\begin{tabular}{lll}
\hline
\noalign{\smallskip}
Telescope & Date              & UT range     \\
\hline
VLT       & 2003 January 7    & 01:23--08:27 \\
WHT       & 2001 December 1/2 & 22:56--06:37 \\
WHT       & 2001 December 2/3 & 21:50--06:59 \\
WHT       & 2001 December 4   & 00:54--04:15 \\
\noalign{\smallskip}
\hline
\end{tabular}
\end{center}
\end{table}

\section{Observations and data reduction}
\label{OBSERVATIONS}

\begin{figure}
\psfig{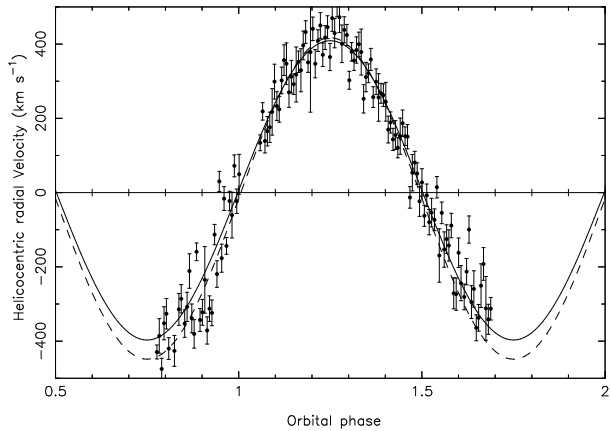}
\caption{ The VLT radial velocity curve of \target. The solid line is the  best
fit sinusoid with a semi-amplitude of $K_2$=403\kms and  the dashed line is a
plot with $K_2$ fixed at 432.8\kms \citep{Marsh94}. }
\label{FIG:RVCURVE}
\end{figure}

\subsection{Very Large Telescope}

Time resolved spectroscopic observations of \target\ were obtained on Jan 7
2003  using ESO's Very Large Telescope (VLT) Unit Telescope 1 equipped with the
Focal Reducer Low  Dispersion Spectrograph FORS1. A log of the observations are 
given in Table\,1. The data were taken with the 600V grism with an order-sorting 
filter (GC435+31) centred at 6270\AA\ and integration times of 120\,s.  
To maximise spectrophotometric accuracy a wide slit was used, and as a result 
the spectral resolution was determined by the seeing ($<$0.8\arcsec for the 
entire  night).  We obtained 32 spectra with a slit width of 2.5\arcsec and and
96 spectra with a slit width of 2.0\arcsec.   The wavelength coverage and 
dispersion depends on the slit width, and we measured from the arc lines  
coverages and dispersions of 4933-7329\AA at 1.17 \AA per pixel for the 
2.5\arcsec slit and 4452-6794\AA at 1.15 \AA per pixel for the 2.0\arcsec slit.
The  corresponding effective resolutions set by the seeing were about 4.7 \AA 
and 4.6 \AA for the 2.5\arcsec and 2.0\arcsec wide slits, respectively.
Although the atmospheric dispersion corrector was used, in order to minimise
atmospheric dispersion  along the slit when observations were taken at high
airmass,  the slit was aligned to a position angle of 131.7\,degrees East of
North. The crowded  field of view  of \target\ is such that we were able to
center a non-variable  comparison star  (38\arcsec North 42\arcsec West of \target) 
on  the slit for slit light-loss corrections. Finally, since the observing 
conditions were excellent, the spectrophotometric  flux standard HZ2 \citep{Oke74} 
was observed.

\begin{figure*}
\psfig{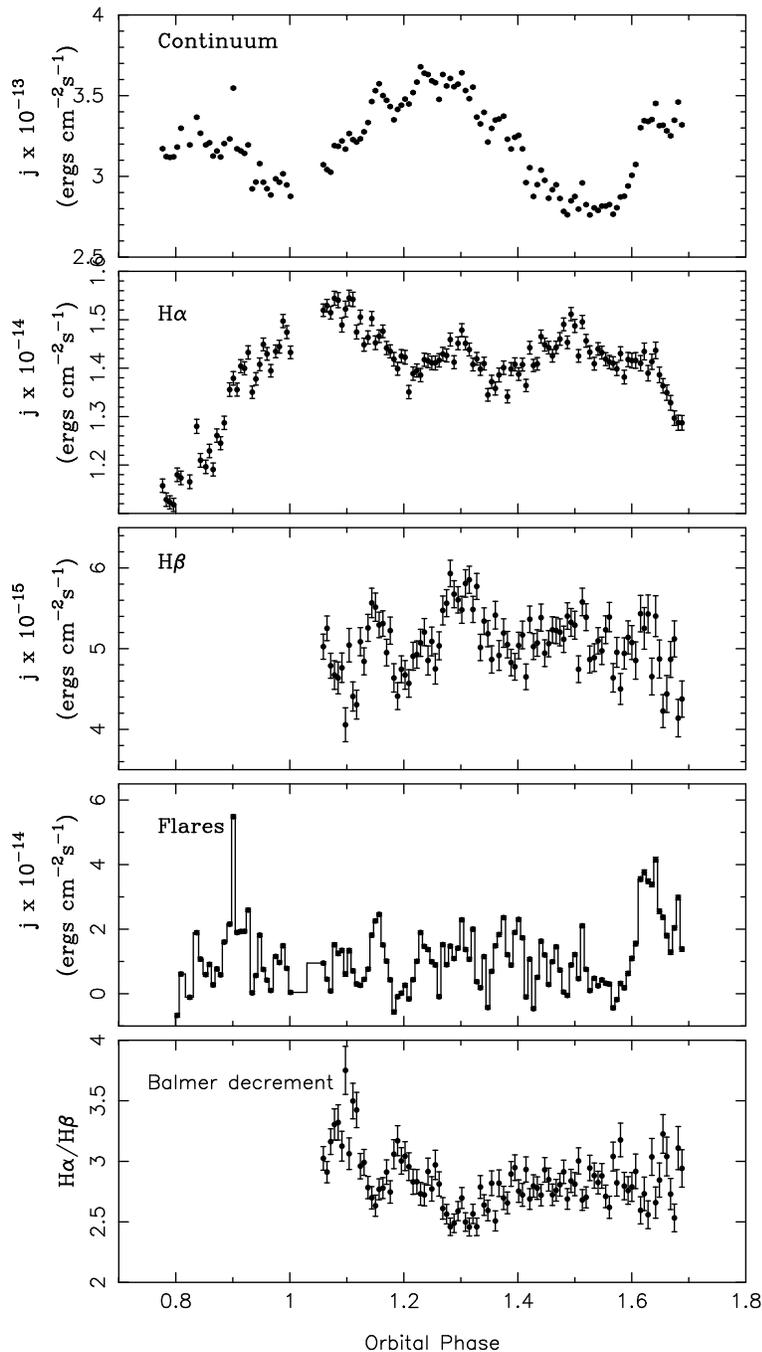}
\caption{VLT continuum and Balmer emission line lightcurves of \target. From
top to bottom: the lightcurve of the continuum flux from 4500--6700 excluding
the emission lines (see text): the H$\alpha$ and H$\beta$ emission line
continuum subtracted lightcurves respectively; the   lightcurve of the flares
obtained by subtracting a fit to the lower-envelope of the continuum
lightcurve  and the H$\alpha$/H$\beta$ continuum subtracted, integrated-flux
ratio lightcurve. 
 The error bars for the continum lightcurve are smaller than
the symbol.} 
\label{FIG:VLT_LCURVES}
\end{figure*}

The data were debiased and corrected for the pixel-to-pixel variations using
the bias frames and tungsten lamp flat-fields respectively,  taken as part of
the daytime calibrations.  The sky was subtracted by fitting second-order 
polynomials in the spatial direction to the sky regions on either side  of the
object. The spectra of \target\ and the comparison star  were then optimally
extracted \citep{Horne86} with  arc spectra extracted at the same location on
the detector as the targets. The wavelength scale was determined through a
fourth-order polynomial fit to more than 25 arc lines  giving a root-mean
square error of $<0.5$\AA\ for all the spectra. Some residual flexure is to be
expected, and the additional  use of a wide slit means that there is some
uncertainty due to the position of the centre of light of the target.  The
latter term means that it is impossible to use the night sky lines to correct
for this, so the  comparison star's H$\alpha$ absorption line was used.  The
relative velocity shift of each comparison star spectrum with respect to
the first spectrum was computed  using the method of cross-correlation.

The velocity shifts were then applied to \target\ and the comparison star
spectra. Finally the velocity shift of the first comparison star, computed
using the position of the H$\alpha$ absorption line, was  added to
the \target\ velocities.

We corrected the spectra for the instrumental response and determined the
slit-losses to obtain absolute flux calibrated spectra. A third order
polynomial fit to the continuum of the  spectrophotometric standard was used to
remove the large-scale variation of the instrumental response and the slit-loss
corrections were calculated using the comparison  star spectra.
Finally we use the colour excess of $E(B-V)$=0.35$\pm$0.02  \citep{Wu83} and
the extinction law of \citet{Seaton79}  to correct the spectra of \target\ for
interstellar reddening  assuming $A_{\rm V}/E(B-V)$=3.1. This is the starting
point for the analysis of the spectra  presented hereafter.

\subsection{William Herschel and Jacobus Kapteyn Telescopes}

Spectrophotometric observations of \target\ were also obtained with the William
Herschel Telescope (WHT) on 2001 December 1--4 using the ISIS dual-beam
spectrograph in single red arm mode to maximise throughput. The observations
used the TEK4 CCD and R158R grating yielding a dispersion of 2.9\,\AA\ per
unbinned pixel. All observations were either unbinned or we used $2\times2$
binning. The exposure time used was 120\,sec. A wide slit of 4\,arcsec rotated
to include a comparison star (which was not the same as the comparison star 
used at the VLT) was used to ensure spectrophotometric accuracy,
so the actual spectral resolution was set by the seeing. This was
1.9--2.4\,arcsec on the first night, 1.0--1.7\,arcsec on the second night, and
about 1.0\,arcsec on the third night, corresponding to effective resolutions of
15--20\,\AA, 8--14\,\AA, and about 8\,\AA\ respectively.
A log of the observations are given in Table\,1.

All images were debiased and flat-fielded using standard 
{\sc iraf}\footnote{{\sc iraf} is distributed by the National Optical Astronomy
Observatories, which are operated by the Association of Universities for
Research in Astronomy, Inc., under cooperative agreement with the National
Science Foundation.} techniques. One-dimensional spectra were then extracted of
both \target\ and an on-slit comparison star using the optimal extraction
method of \citet{Horne86}.  Variations in transparency and wavelength dependent
slit-losses were corrected by fitting a smooth function to the ratio of the
individual comparison spectra to their average. Wavelength calibration was done
by interpolating from a series of CuNe+CuAr arc spectra obtained at intervals
of approximately an hour during each night.  As the comparison star did not
have strong enough absorption lines, further correction for flexure was not performed.
Since the velocity resolution is low anyway (350--900\,km\,s$^{-1}$) we decided
to discard the detailed velocity information and use the H$\alpha$ emission
in \target\ to calculate wavelength corrections, and simply work with
integrated line fluxes. All three nights were non-photometric, so no attempt
was made to apply absolute flux calibration and all fluxes are expressed
relative units.

Simultaneous photometry was obtained with the SITe2 CCD camera on the
Jacobus Kapteyn Telescope (JKT).  An $R$ band filter was used with 60\,s
exposures.  Standard {\sc iraf} data reduction techniques were used to
process the images, and differential photometry was performed using the
optimal photometry algorithm of \citet{Naylor98}.

\begin{figure}
\hspace{-5mm}
\psfig{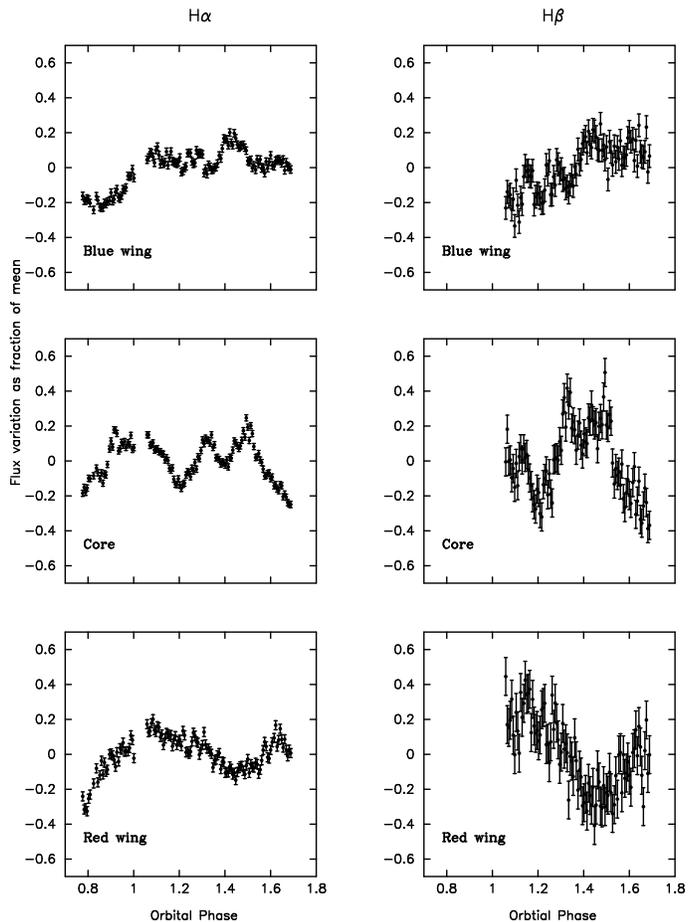}
\caption{ VLT balmer emission line lightcurves of \target. From top to bottom;
the blue-wing, core and red-wing emission line continuum-  subtracted
lightcurves. The left and right panels are for  H$\alpha$ and H$\beta$
respectively. }
\label{FIG:BCURVES}
\end{figure}

\begin{figure}
\psfig{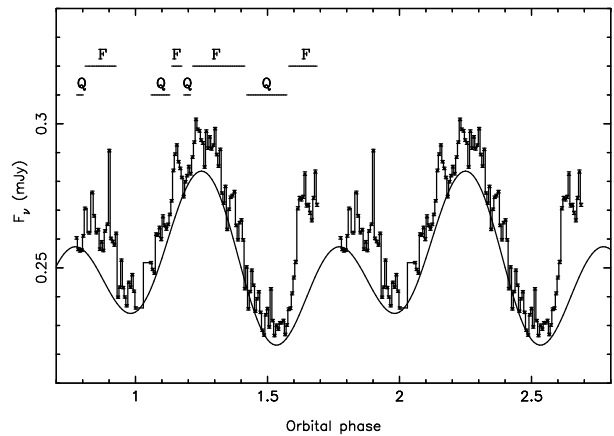}
\caption{
The VLT continuum lightcurve of \target\ with the positions of the  quiet-state
(Q) and flare-state (F) marked. The solid curve is a fit to the lower-envelope
and represents the light from the secondary star and steady accretion disc. For
clarity the data  have been plotted over two orbital cycles.
The error bars are smaller than symbol size }.
\label{FIG:FLAREPOS}
\end{figure}

\begin{figure*}
\psfig{angle=90,height=10.0cm,file=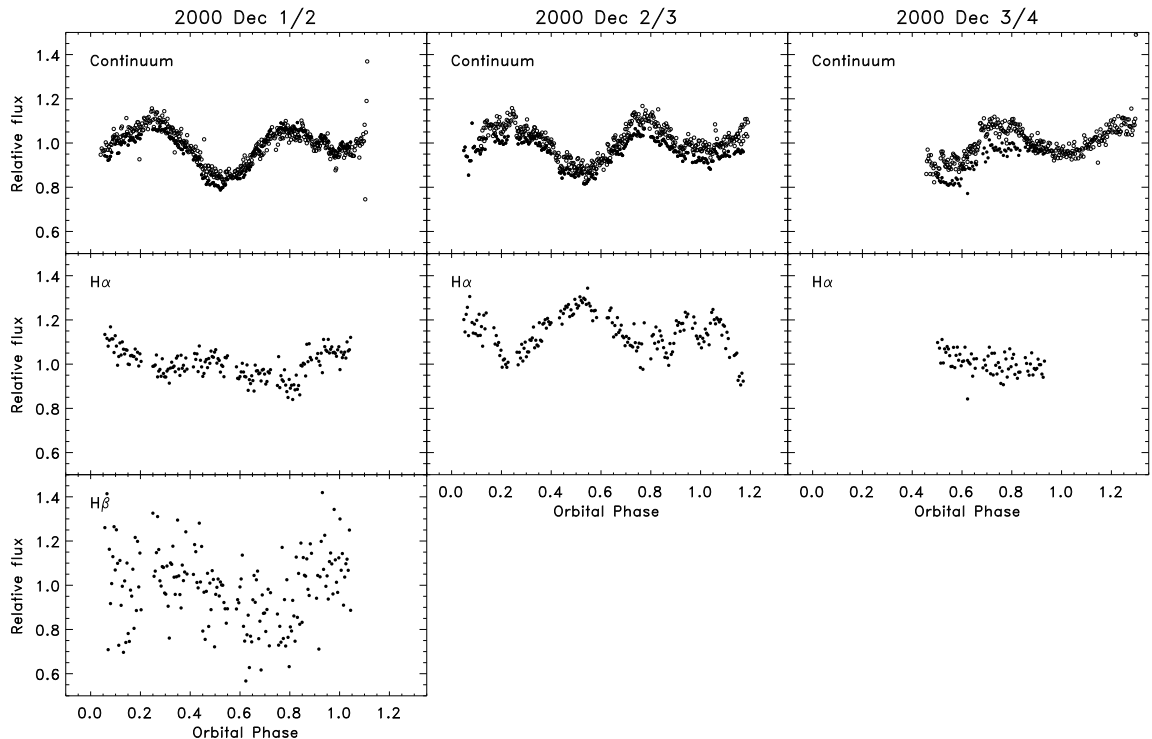}
\caption{WHT/JKT continuum and Balmer emission line lightcurves of A0620--00. 
All fluxes are relative to the median value on the first night.  In the
continuum panels, open circles are JKT $R$ band fluxes, solid circles are WHT
4500--6700\,\AA\ continuum fluxes (with the emission lines masked out). }
\label{FIG:WHT_LCURVES}
\end{figure*}

\section{The secondary star's radial velocity curve}
\label{K2}

Since the use of a wide slit adds uncertainty to the  wavelength calibration,
we only use the VLT to  determine the radial velocity curve, as no appropriate 
correction could be performed for the WHT data.
We took the VLT data and  used the cross correlation method of \citet{Tonry79}
to determine radial velocities of absorption lines pertaining to the secondary 
star.  Prior to cross-correlation the spectra were interpolated onto a constant
velocity  scale (60 \kms\,pixel$^{-1}$) and  normalised by fitting a spline
function to the continuum. Only regions of the spectrum devoid of emission
lines  (5000--5820\AA\ and 5900--6500\AA) were used in the analysis. The Balmer
emission lines (H$\beta$ and H$\alpha$), HeI (5875\AA) and interstellar NaI\,D
lines were masked out so that they did not affect the cross-correlation
process.

We use the comparison star as a template, since its spectrum contains features
of a late-type K-star. The resulting radial velocity curve was then fitted with
a sinusoid with the orbital period fixed at $P_{\rm orb}$=0.323016\,d. We
determine  the radial velocity semi-amplitude $K_2$ to be  403.0$\pm$4.5\kms, 
the systemic velocity $\gamma$=10$\pm$2\kms and the time at orbital phase 0.0, 
$T_0$ to be 2452646.6365$\pm$0.0005\,d; phase 0.0 corresponds to the inferior 
conjunction of the secondary star. The reduced $\chi^2$ of the fit was 2.7 and the 
uncertainties quoted are 1-$\sigma$ and have been rescaled so that the  reduced
$\chi^2$ of the fit is 1. It should be  noted that since we used the position
of the  comparison star's H$\alpha$ absorption line to determine the velocity 
zero point  of the instrumental flexure, the radial velocities are relative to
the comparison star. Furthermore, given the uncertainties in wavelength calibration 
due to the large slit width, the value for $\gamma$ should not be taken at face value.
The radial velocity curve and sinusoidal fits are shown in Figure\,\ref{FIG:RVCURVE}. 
The scatter in the radial velocity curve is caused by the uncertain wavelength
calibration, due to the use of a wide slit. We also used a template star spectrum 
obtained by shifting all the spectra of \target\ into the rest frame of the  secondary
star, using the radial velocity solution derived  by  \citet{Marsh94}. The
results obtained are the same as those obtained using the comparison star as a
template. The amplitude we obtain for $K_2$ is less than that obtained by
\citet{Marsh94}. This is because our coverage of the minimum in the radial
velocity curve at phase 0.25 is not complete and so the derived value for 
$K_2$ is biased. By fitting the radial velocity curve with $K_2$ fixed at 432\kms 
\citep{Marsh94} we obtain $T_0$=2452646.6361$\pm$0.0005\,d. It should be noted 
that the $T_0$ we obtain
is different by 0.43 phase to  that quoted by \citet{Gelino01}, even though the
nominal definition of orbital  phase 0.0 adopted is the same.  This may have
arisen from the potential ambiguity of defining phase 0.0 photometrically, as
the minima at phase 0.0 and 0.5 can be almost the same depth and hence can be
confused.  No such ambiguity exists with our spectroscopic ephemerides.

\section{The flare lightcurves}
\label{LIGHTCURVES}

In Figures\,\ref{FIG:VLT_LCURVES} and \ref{FIG:BCURVES} we present
the VLT continuum and emission line
lightcurves as a function of orbital phase.  The continuum lightcurve was
obtained by summing the flux across the 5000--5820\AA\ and 5900--6500\AA\
regions.  The continuum lightcurve exhibits rapid short-term variations as
seen  previously by \citet{Haswell92}, \citet{Zurita03} and \citet{Hynes03}.  
To determine the flare lightcurve we subtract a fit to the lower envelope 
of the lightcurve. We use an iterative rejection scheme to fit the lower
envelope, similar to \citet{Zurita03} and \citet{Hynes03}. 
We reject points more than 2-$\sigma $above the fit, then refit, repeating 
the procedure until no new points are rejected.
Most noticeable are the brief  flares at orbital phase 1.15 and 1.2  which have
a rise time of $\sim$15\,min and amplitudes of 3 and 12 percent  respectively. 
A larger flare is also seen at phase 0.6 which has a rise time of $\sim$30\,min 
and an amplitude of $\sim$20 percent. In Figure\,\ref{FIG:FLAREPOS} we have marked the
beginning and end of  the strongest most significant flare events and also 
the periods when flares are not seen. 
 
To extract the Balmer line fluxes, the continuum background was removed by
fitting and subtracting a smooth polynomial fit to the continuum regions 
around the lines before the line flux was integrated.  The Balmer line
lightcurves show significant variations, scatters of  3.8 and 7.6 percent for 
the H$\alpha$ and H$\beta$ lightcurves respectively. The  H$\beta$ and
H$\alpha$ lightcurves are strongly correlated.  Figure\,\ref{FIG:BCURVES} shows
the H$\alpha$ and H$\beta$ emission line  wings and core
determined by integrating the line flux. 

The blue, core, and red wing fluxes were calculated by integrating the 
emission line over the velocity ranges  --2500 to --500\kms, -500 to 500\kms
and 500 to 2000 \kms  respectively. One can clearly see that the  H$\alpha$ and
H$\beta$ emission line  wings and core are correlated with each other.   The
Balmer line lightcurves show some correlations with the flare lightcurve
(Figure\,\ref{FIG:VLT_LCURVES}).   This is clearest  for the flare event at
phase 1.15. There is no noticeable  participation from H$\alpha$, although the
correlation  is quite strong in H$\beta$.   However, it could be that there is
an underlying source of variable  contamination in the H$\alpha$ lightcurves
such as from the disc  (see section\,\ref{DOPPLER}), that prevents a complete
correlation with the  continuum  lightcurve.  The fractional variability in
H$\beta$ is a factor of $\sim$2 larger than in H$\alpha$ (see
Figure\,\ref{FIG:VLT_LCURVES}). In V404\,Cyg the correlation between H$\alpha$
and the continuum is much stronger. Also in  V404\,Cyg the H$\alpha$ emission
line varies considerably \citep{Hynes02} by a factor of  $\sim$2 more than in
\target. However, it could be that the optical  depth in \target\ is larger
than that in  V404\,Cyg, so the emission line flares  are generally weaker, and
H$\beta$ is more pronounced.  Also, the persistent lines could be stronger
relative to the flaring component in \target.

The H$\alpha$/H$\beta$ ratio varies by 8 percent around a mean value
of 2.7, and is inversely correlated with the line fluxes, consistent
with there being a higher optical depth in H$\alpha$ than in H$\beta$.
The ratio H$\alpha$/H$\beta$ is consistent with case B 
recombination  (2.85 for T=10$^4$\,K \citealt{Osterbrock87}) and
so indicates that the Balmer emission lines are optically thin.  
For the flare event at phase 1.15 the Balmer decrement drops from 3.0 to 2.6
during the flare event indicating that either the  flare arises from more
optically thick regions or most likely that the temperature increases  during
the flare (from $\sim$5000\,K to $\sim$30,000 \,K;  \citealt{Osterbrock87}).
The amplitude of the line flare is larger than in the continuum by a factor of
$\sim$3. One might expect that is because the continuum lightcurve is diluted
by light from the secondary star (see section\,\ref{LOW}).  After this
correction the amplitude of the flares in the continuum compared to the line
are similar.  From the data presented here for \target, some, but not all, 
flare events that appear in the continuum lightcurve are also present in the
emission line, similar to what is observed in the continuum and H$\alpha$
lightcurves of V404\,Cyg \citep{Hynes02}.

The WHT lightcurves (Figure\,\ref{FIG:WHT_LCURVES}) exhibit no pronounced flares 
similar to those seen in the VLT data (Figure\,\ref{FIG:VLT_LCURVES}).  
The continuum lightcurves are relatively smooth, with possibly some
weak flares present.  These are not detectable in the H$\alpha$ lightcurves,
however, and the H$\beta$ data, where present, are too noisy to tell.  It may
be that the WHT observations represent a less active epoch; such differences in
flare behaviour between epochs do seem to occur in quiescent SXTs, and can be
very marked (Hynes et al.\ in preparation).

\section{The excess light}

There have been several photometric studies of \target\ in quiescence  which
show a distorted double-humped modulation (\citealt{Shahbaz94};
\citealt{Haswell93}; \citealt{Khruzina95};  \citealt{Leibowitz98} and
\citealt{Gelino01}). The maxima and minima in  the lightcurves vary in height
and depth and  on occasions the maxima have reversed (see Figure\,1 of 
Haswell 1996). Furthermore, it has been reported that the average optical magnitude  
varies on a timescale of a few hundred days. Possible explanations for these 
variations are star-spots on the secondary star (\citealt{Khruzina95}; 
\citealt{Gelino01}) or a persistent superhump \citep{Haswell93}. 

\begin{figure}
\psfig{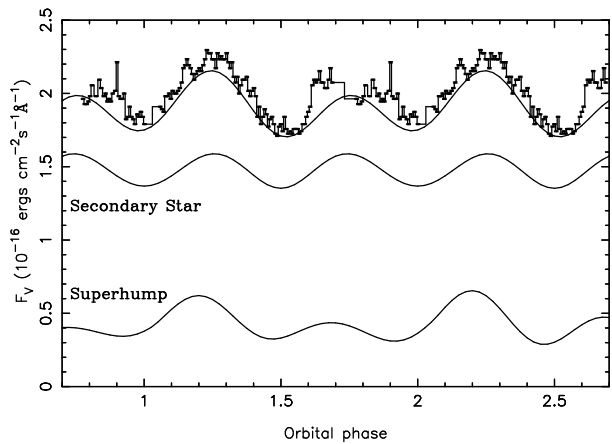}
\caption{ The VLT de-reddened $V$-band lightcurve of \target\ obtained by 
folding the dereddened spectra through the Johnson $V$-band filter response.
The solid curve is a fit to the lower-envelope of the lightcurve  and
represents the light from the secondary star and steady accretion disc. The
secondary star's V-band ellipsoidal modulation scaled to the distance of
\target\ is shown. The excess lightcurve most probably due to the distorted
accretion disc, is obtained by subtracting the  secondary star's ellipsoidal
lightcurve from the observed underlying lightcurve. For clarity the data  have
been plotted over two orbital cycles. 
The error bars are smaller than symbol size.}
\label{FIG:SHUMP}
\end{figure}

To obtain a V-band lightcurve of \target\ from our spectra, we  integrated the 
dereddened spectra with the Johnson $V$-band filter response. The resulting
lightcurve is typical;  we see the classic distorted double-humped modulation
and the maxima have different heights. Our lightcurve resembles the optical 
lightcurves obtained by \citet{Haswell93}, with the brighter maxima  occurring
at phase 0.25. The uncertainty in the reddening adds a 7 percent uncertainty to the 
the mean flux level.We use the X-ray binary model described in \citet{Shahbaz03a}  
with the parameters $q$=0.067, $i$=41$^\circ$,  \bh=11\Msun\, a distance of 
$d$=1164\,pc and a secondary star with an effective temperature of T$_{\rm eff}$=4600\,K 
(\citealt{Marsh94}; \citealt{Gelino01}), to determine the secondary star's
ellipsoidal modulation.  The excess lightcurve is then obtained
by subtracting the secondary star's ellipsoidal modulation from the underlying
continuum of the de-reddened lightcurve  (see Figure\,\ref{FIG:SHUMP}).

To interpret the observed lightcurve we require the secondary star's ellipsoidal 
modulation  plus the steady accretion disc's light with excess continuum emission  
(i.e. superhump light) between orbital phase 0.00 and 0.50. The mean level of the 
excess lightcurve i.e. the photometric veiling, suggests that the disc contributes 
20 percent to the observed flux. This is consistent with the disc fraction estimated 
using the secondary star's absorption lines (in section\,\ref{LOW}),  given that the 
mean level of the ellipsoidal lightcurve is at most only accurate to 20 percent
due to the uncertainties in the distance. Although the exact amplitude and flux of 
the excess lightcurve depends on the model parameters that predict the secondary star's 
ellipsoidal modulation, the result that there is excess light near phase 0.2
remains firm. To determine the colour and possible origin of the excess light we 
compared the data taken around the maximum of the inferred superhump at
orbital phase  $\phi_1 \sim$0.2 (see Figure\,\ref{FIG:SHUMP})
to the data taken one orbital cycle earlier $\phi_2=(1-\phi_1$)low$\sim$0.8. 
Since the secondary star's ellipsoidal modulation contribution will be the same in both 
cases, any difference in the spectral shape must be due to a superhump at $\phi_1$, 
or a starspot depressing  the ellipsoidal maximum at phase $\phi_2$. The continuum shapes 
of the spectra taken at $\phi_1$ and $\phi_2$ are not signifcantly different.
Therefore, our data do not allow us to say if this excess light arises from 
star-spots or the superhump.

\begin{figure}
\psfig{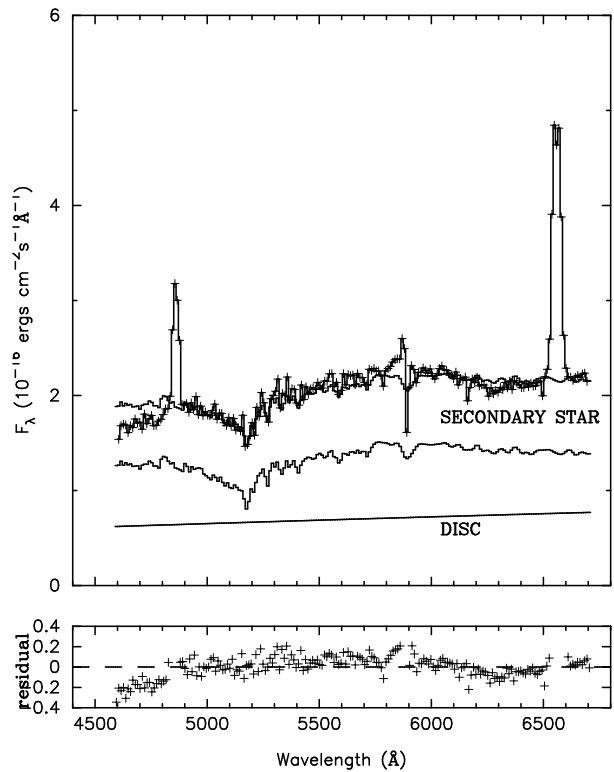}
\caption{ The VLT de-reddened average quiet-state spectrum of \target\
rebinned  for clarity  (histogram with crosses). The best fit (solid line), 
scaled K4 secondary star and power-law accretion disc spectrum are also shown.
The lower panel shows the residuals of the fit. The horizontal dashed line
marks the zero residual line. }
\label{FIG:QSPECTRUM}
\end{figure}

\section{The quiescent spectrum of \target}
\label{LOW}

There have been many studies of \target\ where the  fractional contributions of
the  light from the secondary star and accretion disc have been estimated. By
subtracting the K secondary star spectrum from the broad-band data  of \target\
obtained ten months after the 1975 May outburst,  \citet{Oke77} found the total
V-band light  due to the disc at 5500\AA\ to be $f_{\rm 5500}$=43$\pm$6
percent. He also determined a power-law form  for the disc's light
($F_\lambda \propto \lambda^\beta$) with an index  of -2 . Similarly,  using
the same analysis, \citet{MR86} obtained  $f_{\rm 5500}$=40$\pm$10 percent and
$\beta$=-2.5$\pm$1.0 for the disc's light in 1985. Further observations by
\citet{MHR95} showed that the disc's fractional contribution at 5225\AA\ is
roughly constant at $f_{\rm 5225}\sim$42 percent during 1985--1990.  Using high
resolution spectroscopy, \citet{Marsh94} find that the disc  contributes
6$\pm$3 percent  at H$\alpha$ and 17$\pm$3 percent at H$\beta$ in 1991/1992.
As pointed out by \citet{MHR95} the difference in the disc's fractional contribution 
determined by \citet{Marsh94} and \citet{MHR95} is probably due to the choice
of the template star.

We model the average quiet-state spectrum of \target\ in terms of a template
secondary star and non-variable accretion disc spectrum. We use a K4 spectrum
(HD154712) taken from the Gunn-Stryker Atlas  \citep{Gunn83} to represent the
secondary star and a power-law spectrum of the form  $F_\lambda \propto
\lambda^\beta$ to represent the light from the  accretion  disc.  Since the
spectral resolution of the \target\ spectrum is higher than the template stars
in the Gunn-Stryker Atlas, the spectrum of  \target\ was rebinned to the
spectral dispersion of the template star (10\,\AA\,pixel$^{-1}$).
Figure\,\ref{FIG:QSPECTRUM} shows the results of a spectral synthesis of the
continuum emission.  We find that the accretion disc's light has a power-law
index of 0.60$\pm$0.3 and contributes 37$\pm$13 percent to
the observed continuum light at 5500\,\AA. From the residual of the fit it 
can be seen that a power-law model for the disc predicts more flux short-ward of 
5000\AA, so a power-law function does not best describe the disc's light.
If we model the spectrum of the disc with a blackbody function, we obtain a 
significantly better fit (at the 99 percent level), where the disc contributes 
58$\pm$16 percent to the continuum flux at 5500\AA\ and the disc has a blackbody 
temperature of 4600$\pm$100\,K and a radius of 0.50$\pm$0.06\Rsun. Using a K3 star 
in order to match the blue end of the spectrum does not give a better fit (at the 
99.99 percent confidence level.) All the uncertainties quoted are 1-$\sigma$ and have 
been rescaled so that the  $\chi_{\nu}^{2}$ of the fit is 1.

Although our determination of the fractional contribution of the accretion
disc's  light to that observed is consistent with the findings of
\citet{Oke77}, \citet{MR86} and \citet{MHR95}, the form of power-law
description of the disc's light  is not.  However, it should be noted that the
non-variable accretion disc light is a composite of light from the steady-state
accretion disc plus the excess light from star-spots or a late-superhump, which 
most probably have very different spectral shapes. For the case where the excess 
light is due to a late-superhump, the late-superhump modulation may have its origin 
in the outer regions of the disc in the changing stream-disc interaction 
(Rolfe, Haswell \& Pattterson 2001), the spectrum of the late-superhump is
expected to be different compared to the steady-state disc. 
Although the tidally heated stream-disc impact region will be
hotter than the rest of the outer disc, it is not obvious how much hotter
this will be compared to the inner regions of the disc, where viscous stresses
due to differential rotation are much higher. Thus it is difficult to 
estimate the spectrum of the late-superhump without detailed computations.
Therefore, it is no surprise that the disc's light and spectrum are observed to change 
with time,  as it just reflects the variable behaviour of the accretion disc.

\section{The flare spectrum}
\label{FLARE}

\begin{figure}
\psfig{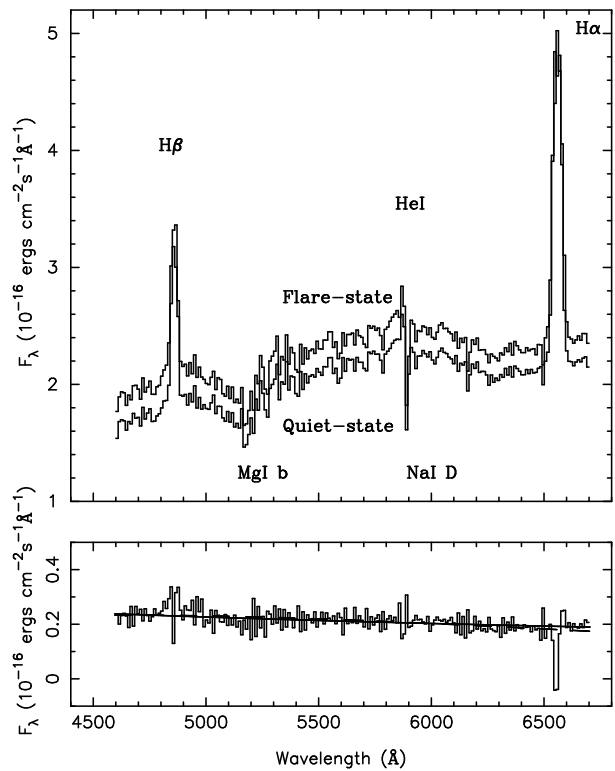}
\caption{ Top panel. The average VLT dereddened quiet-state (Q) and flare-state
(F)  spectra of \target. The spectra have been re-binned for clarity. Bottom
panel. Spectrum of the flare obtained by subtracting the quiet-state  spectrum
from the flare-state spectrum. The solid line shows a continuum  fit using a 
hydrogen slab in LTE at 12000\,K and a baryon column density of  10$^{22}$
nucleons\,cm$^{-3}$.}
\label{FIG:FSPECTRUM}
\end{figure}

Assuming that the light produced by the flares is simply added to the quiescent
spectrum which contains the secondary star and  light from the accretion disc
(assumed to be non-variable),  the difference between flare-state spectra and 
the quiet-state spectra yields an estimate for the spectrum of the  flares. The
actual portions of the lightcurves used to determine the  flare-state and
quiet-state  spectra are marked in   Figure\,\ref{FIG:FLAREPOS} and were
selected after subtracting the secondary star's ellipsoidal modulation.
Figure\,\ref{FIG:FSPECTRUM} shows the resulting spectrum, which has been binned
for clarity. 

We compared the average flare spectrum taken during the beginning and end  of
the night. They were found to be the same to within 1.5 percent. Before the
subtraction we shift the spectra into the rest frame of the secondary star, so
that the features from the secondary are removed cleanly.  
The R-band mag
varies on a timescale of a  few hundred days with an  amplitude of 0.3 mag
\citep{Leibowitz98}. However, since this timescale is much longer than the
orbital period it is safe to assume that the lightcurve of the  superhump of
excess light does not change shape  over the length of the  orbital period.

The flare spectrum has a relatively flat continuum suggesting  that a high
temperature model is needed to fit the data. We attempt to fit the flare
spectrum with different models. A fit using a blackbody ($\chi^{2}_{\nu}$=3.93)
gives  a temperature of 5900$\pm$200\,K  and a radius  0.14$\pm$0.10\Rsun\ (90
percent confidence). A power-law fit  ($\chi^{2}_{\nu}$=3.85) of the form 
$F_\lambda \propto \lambda^\beta$ has an index of -0.60$\pm$0.20,  or  
$F_\nu \propto \nu^{-1.40\pm0.20}$ (90 percent confidence).

A more realistic model is a continuum emission spectrum  of an LTE slab of
hydrogen. Using the synthetic photometry SYNPHOT package ({\sc iraf/stsdas})
to compute the LTE models we fit the continuum regions
to estimate the temperature, radius  and baryon column density of the slab. We
find that the $\chi^2$ surface has a broad minimum at
$\chi^{2}_{\nu}$ of $\sim$4. The requirement that the flare region is smaller than the
area of the accretion disc (0.5\,$R_{\rm L1}$; \citealt{Marsh94})  adds a weak
constraint, only ruling out temperatures less than $\sim$5000\,K. 

\begin{figure}
\psfig{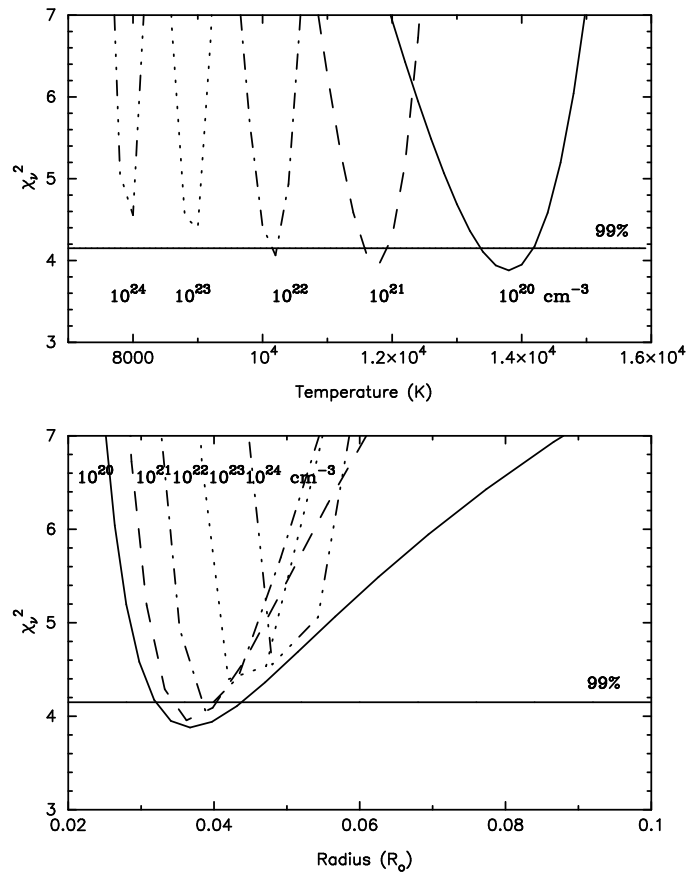}
\caption{
The results of LTE model fits to the VLT flare spectrum.  The top and bottom
panels show the reduced $\chi^{2}$ versus temperature  and radius fits
respectively. In each panel, the results for different baryon column
densities are shown.  The horizontal line marks for 99 percent confidence
level.
}
\label{FIG:CHI}
\end{figure}

We constrain the temperature and equivalent radius of the optically thin LTE 
slab to lie in the 
range 10000--14200\,K and 0.032--0.044 \Rsun\ (99 percent confidence) for 
a baryon column density in the range 10$^{20}$--10$^{24}$ nucleons\,cm$^{-3}$
(see Figure\,\ref{FIG:CHI}). The emission covers 0.05--0.08 percent 
(99 percent confidence) of the accretion disc's projected surface area 
($q$=0.067, $i$=41$^{\circ}$ and $M_{\rm 1}$=11\Msun; \citealt{Marsh94} 
and \citealt{Gelino01}). Tight constraints on the
parameters of interest  cannot be placed because of the correlations between
the temperature and column density; a high temperature and low column density
model gives  the same $\chi^2$ as a low temperature and high column density
model. Although this can only be done by using the Balmer emission line 
fluxes, for \target\ this is very difficult as it is clear that the Balmer
emission lines are contaminated by the emission from the bright-spot  (see
section\,\ref{DOPPLER}).

\subsection{A comparison with AE\,Aqr}

AE\,Aqr is an unusual cataclysmic variable in that it exhibits flaring 
behaviour which can be described in the framework of a magnetic propeller 
that throws out gas out of the binary system \citep{Wynn97}.
The flares are thought to arise from collisions between high-density
regions in the material expelled from the system after interaction with the
rapidly rotating magnetosphere of the white dwarf \citep{Pearson03}.
The spectrum of the flares in AE\,Aqr can be described by a optically thin gas
with a  temperature of 8000--12,000\,K \citep{Welsh93}.
The spectrum of the flares seen in \target\ are described by 
optically thin gas with a similar temperature, it seems unlikely that the same
mechanism producing the flares in AE\,Aqr operates in the 
SXTs. [Although the temperature of the large flares in V404\,Cyg have 
been estimated to be $\sim$8000\,K, the uncertainties 
should be noted \citep{Shahbaz03b}.]

\section{The Doppler maps}
\label{DOPPLER}

Doppler tomography is a powerful tool in the study of accretion disc kinematics
and emission properties  \citep{Marsh88}. This analysis method is now a widespread 
procedure in the study of emission-line profiles in both cataclysmic variables  and 
low-mass X-ray binaries, providing a quantitative mapping of optically thin emission 
regions in the velocity space.  When intrinsic Doppler broadening by bulk flow is 
present, the tomograms provide a concise and convenient form of displaying 
phase-resolved line profiles.  However, there are several assumptions inherrant in 
Doppler tomography that should be noted; it is assumed that the emission is fixed 
in the co-rotating frame, the emission does not vary over the binary orbit and 
the motion is in the orbital plane. If these  assumptions are not valid  then 
Doppler tomography can produce artefacts \citep{Foulkes04}.
 
The Balmer line profiles show the doubled peaked structure typical of an
accretion disc, and periodic changes in the strength of the two emission peaks 
which are attributed to S-wave emission components  (see Figure\,\ref{FIG:DOPMAPS}).  
The trailed spectra show two S-wave components. Prior to constructing the
Doppler images, the spectra were re-binned onto a constant velocity scale (60
\kms\,pixel$^{-1}$) and then continuum subtracted using  a spline  fit to the
continuum regions close to the Balmer lines.  When computing the Doppler maps, 
we use the systemic velocity determined from the  secondary star's absorption
radial velocity curve \citep{Marsh94}.

\subsection{The bright-spot}
\label{BSPOT}

The H$\alpha$ and H$\beta$ Doppler images show a broad ring background which
is the emission from the accretion disc extending to high velocities.  The
images also show regions of entended enhanced emission at the position of the 
stream/disc impact point and extended emission from the opposite side of the
disc. 

\begin{figure*}
\psfig{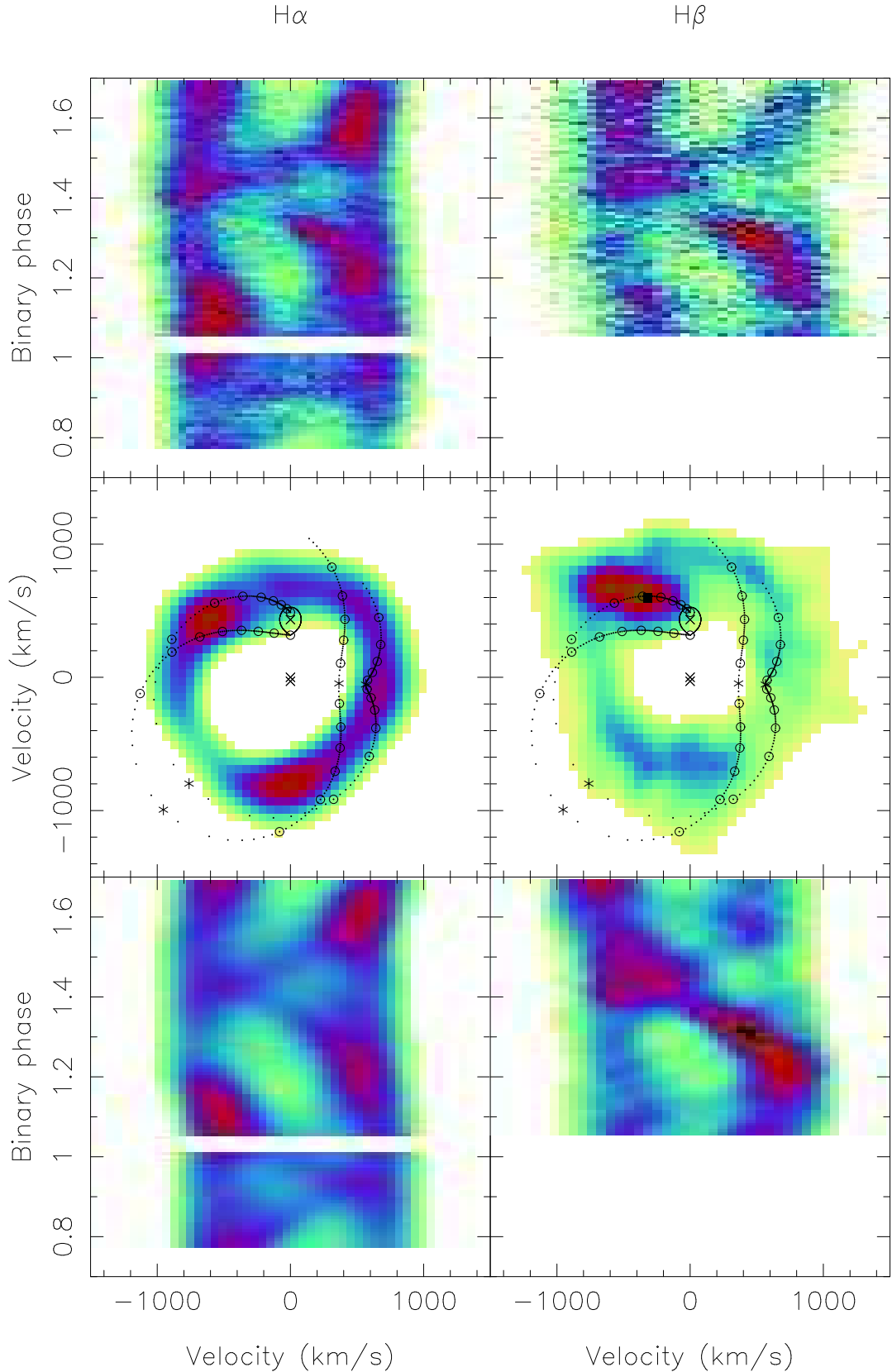} 
\caption{The Doppler images of H$\alpha$ (left panel) and H$\beta$ (right
panel). From the top to bottom,  the VLT data, the Doppler image and the fitted
data. The ``Doppler ghost'' of the  secondary star is shown along with the
system's  centre of mass and the  velocities of the compact object and
secondary star (X).  The two curves show the theoretical gas stream path from
the inner Lagrangian point and  Kepler velocity along the gas stream for
$q$=0.067 and $K_2$=433\kms.  The asterisks mark turning points in distance
along the gas stream relative  to the compact object. Circles have been plotted
every 0.1\,$R_{\rm L1}$ and dots every  0.01\,$R_{\rm L1}$ along the streams. }
\label{FIG:DOPMAPS}
\end{figure*}

In Figure\,\ref{FIG:DOPMAPS} we show paths for the gas stream and the Kepler 
velocity of the disc  along the gas stream for $q$=0.067 and $K_2$=433\kms. 
The bright-spot does not lie on either predicted path for the  gas stream,
but occurs half-way between them as is observed  in U\,Gem \citep{Marsh90}.
As explained by \citet{Marsh94} the most probable explanation for this  is
that we are seeing gas around  the bright-spot after it has passed through
the shock at the edge of the accretion disc. Assuming that the velocity after
the shock is a combination of  the velocity of the gas stream and accretion
disc, we can  estimate the radius of the post-shock emission, which most
probably  corresponds to the accretion disc radius. From the H$\alpha$ and
H$\beta$  maps we deduce a radius of   $R_{\rm disc}$=0.55$\pm$0.05\RL 
(where \RL is the distance the the  inner Lagrangian point),   similar to
the  values  determined previously,  $R_{\rm disc}\sim$0.5\RL 
\citet{Marsh94}. 

The H$\alpha$ data suggest an anti-correlation with the continuum lightcurves,
in both the WHT and VLT datasets.  Such an effect would, of course, be expected
in the equivalent width, but not in the fluxes. Given that the effect repeats
between the two independent datasets, this is unlikely an artifact, and
probably rather indicates a real modulation in the H$\alpha$ flux.  Such an
effect can be explained if it originates from a more  optically thick region 
for which we see a varying projected area.  In this case, a double-humped
modulation, analogous to ellipsoidal modulations, is expected.  The most likely
region to be responsible for this is the accretion stream impact and/or
overflow, since the H$\alpha$ and H$\beta$ Doppler tomograms show that this
region is much more prominent in H$\beta$ relative to the disc than in
H$\alpha$, as expected if the H$\alpha$ emission arises from more
optically thick regions than the disc.
Furthermore, we would expect emission to be out of phase with the
ellipsoidal modulation.  The phase of maximum light would depend on the disc
radius and how extended the stream impact region is, but up to a shift of 0.2
or so in phase relative to the ellipsoidal modulations is not implausible.

\section{Discussion}

\subsection{An eccentric disc?}
\label{EDISC}

Superhumps are optical modulations that arise when an accretion disc  expands
to the resonance radius, becomes elliptical and is forced to  precess
\citep{Whitehurst91}. It is believed that enhanced viscous dissipation in the
eccentric disc due to the tidal stress exerted by the  secondary  star is the
source of light observed in the superhump lightcurves. Such an eccentric disc
which gives rise to asymmetric emission line  profiles, has been successfully
used to explain the asymmetric line profiles  in OY\,Car \citep{Hessman92}, 
Z\,Cha (\citealt{Vogt82}, \citealt{Honey88}) and AM\,CVn \citep{Patterson93}.

The Doppler maps of \target\ presented here show significant extended emission,
strongest in H$\alpha$, with similar velocities to the stream/disc impact
point, but from the opposite side of the disc between phase 0.25 and 0.50 (see
Figure\,\ref{FIG:DOPMAPS}).  This crescent-like structure is reminiscent of the
feature seen in the Doppler maps of XTE\,J2123-058 in quiescence
\citep{Casares02}.  Also, the orbital average H$\alpha$ profiles of
XTE\,J1118+480 near quiescence show significant velocity variations and can be
interpreted as evidence for an eccentric precessing disc \citep{Zurita02}. 
Thus the most likely explanation for the crescent structure seen in the Doppler
maps of \target\ presented here, is that it is due to an elliptical accretion
disc.  However, it should be noted that such structure is not seen in the
Doppler maps determined by \citet{Marsh94}.

It is interesting to see what the Doppler map of an eccentric disc would look
like.  \citet{Smith99} have attempted to compute such a map and find that the
Doppler map of an eccentric disc is shifted with respect to the system's 
centre of mass with asymmetric enhancements in intensity around the edge  of
wthe disc. 
More recently, \citet{Foulkes04} have performed detailed 2-D smoothed particle
hydrodymanical (SPH) simulations of an eccentric precessing disc in a 
binary system.
The synthetic trailed spectra show two distinct non-sinusoidal 
S-wave features which evolve with the disc precession phase which are
produced by the gas stream and the disc gas at the stream-disc impact shock.
These emission regions are not fixed in the co-rotating frame and 
vary with time over the binary orbit.
To show the effects of using data that violate the assumptions inherrant in 
Dopper  tomography, the authors produce maximum entropy Doppler maps of  the
SPH trailed  spectra. They find that an artefact of Doppler tomography is that 
the stream impact emission is reconstructed in velocity space as a  broad
emission  region between the gas stream ballistic path and the velocity of a
circular Keplerian disc along the ballistic stream trajectory i.e. the stream's
``Kepler shadow'' (see Figure\,9 of \citealt{Foulkes04}).

In both the H$\alpha$ and H$\beta$ trailed spectra of \target, characteristic 
rapidly turning red-blue-red non-sinusoidal S-wave features between 
orbital phases 0.2 and 0.6 are seen, and look very similar to the simulated SPH
trailed spectra of a precessing disc.  
The reconstructed data
between phase 0.8 and 1.0 indicate that the real data are not well 
reproduced by sum of sinusoids that vary on the orbital period. If there is an
eccentric precessing disc, as there must be if there is a superhump, the 
emission regions are not fixed in the co-rotating frame, it is not surprising
that Doppler tomography fails.
Furthermore, the broad
emission near the stream/disc impact region can be interpreted as the 
artefact of Doppler tomography, which blurs the emission produced 
by the gas stream and the disc gas at the stream-disc impact region.
Thus qualitatively our data suggests the presence of a precessing disc in
\target.

Given the mass ratio of \target, the 3:1 resonance is the dominant tidal
instability and lies at $R_{\rm tidal}$=0.66\RL.  Although we derive a disc 
radius less than this value (see section\,\ref{BSPOT}), it should be noted
that we actually determine the post-shock radius of the gas stream as it
impacts the accretion disc, which most probably lies close to but not on the
edge of  the accretion disc.  
Furthermore, \citet{Foulkes04}, \citet{Hessman92} and \citet{Rolfe01} 
show that the effective radius of the
disc as measured by the stream-disc impact region changes dramatically with
superhump phase. Thus the method of Doppler tomography will estimate a mean
value for the disc's radius.
Therefore, it is possible that the radius of
the accretion disc extends up to the tidal radius and so the asymmetric
brightness  distribution in the accretion disc we observed can be explained
by the presence of  a tidally distorted accretion disc.

\subsection{The mechanism for the flare emission}

\citet{Narayan96} and \citet{Narayan97} have proposed  an   accretion flow
model to explain the observations of  quiescent black hole X-ray transients.
The accretion disc has two components, an inner hot advection-dominated
accretion flow  (ADAF) that extends from the black hole horizon to a transition
radius  and a thin accretion disc that extends from  the transition disc to the
edge of the accretion disc.   Interactions between the hot inner ADAF and the
cool, outer thin disc,  at or near the transition radius could be a  source of
quasi-periodic variability.  The ADAF flow requires electrons in the gas to
cool via synchrotron,  bremsstrahlung, and inverse Compton processes which
predict the form of the  spectrum from radio to hard X-rays. 

It has been suggested that the flares observed in \target\ and other quiescent
X-ray transients arise from the transition radius (\citealt{Zurita03};
\citealt{Shahbaz03b}). Indeed, the 0.78\,mHz feature detected in V404\,Cyg has
been used to determine the transition radius, assuming that the periodicity
represents the Keplerian period at the transition between the thin and
advective disc regions \citep{Shahbaz03b}, and a possible low frequency break
in the power spectrum of \target\ may have a related origin \citep{Hynes03}. In
the original ADAF models the optical flux is produced by synchrotron emission. 
As pointed out in \citet{Shahbaz03b}, it is difficult to explain the flare
spectrum in terms of optically thin synchrotron emission, unless the electrons
follow a much steeper power-law electron energy distribution compared to solar
flares; solar and stellar flares have a frequency spectrum with a power-law
index of $\alpha$=--0.5 and an electron energy distribution with a power-law
index of $\sim$--2. 

However, more recently the ADAF models have been questioned, and
substantial modifications proposed.  \citet{Blandford99} emphasised
that the Bernoulli constant of the
gas is positive and hence outflows are possible (as also noted by 
\citealt{Narayan94}).  In the adiabatic inflow outflow solution
(ADIOS) model of \citet{Blandford99} most of the accretion energy that
is released near the black hole is used to drive a wind from the
surface of the accretion disc. Most of the gas that falls onto the
outer edge of the accretion disc is carried by this wind away from the
black hole, with the result that the hole's accretion rate is much
smaller than the disc's accretion rate.
Advective flows are also expected to be convectively unstable, as
also remarked by \citet{Narayan94}.  In these models the central accretion
rate can also be suppressed.  For either of these cases, the effect is
to shift the source of cooling outwards, and the optical synchrotron
emission is reduced (\citealt{Quataert99}; \citealt{Ball01}).  In
these more realistic cases $optical$ flares directly from the inner
flow are less likely.  Optical emission might still arise from the
inner edge of the outer (thin) disc, or from reprocessing of X-ray
variability.  

\subsection{Do the flares arise from the whole disc?}

It was argued by \citet{Hynes02} that since large ($\sim$ few d)
flares in V404\,Cyg involve enhancements of both blue and red wings of
H$\alpha$, they must involve the whole disc.  While these observations
indicate participation by a wide range of azimuths, they leave open
the possibility that only the inner disc is involved.  Using
simultaneous multicolour photometry \citet{Shahbaz03b} determined the
colour of similar large flares in V404\,Cyg. Although the flare parameters
determined are complicated by uncertainties in the interstellar
reddening, the flare temperature was estimated to be $\sim$8000\,K.
Flares on timescales similar to those present in \target\ (i.e. tens
of minutes) were also observed, but no physical parameters could be
determined given the large uncertainties in the colour. The large
($\sim$ few hrs) flares were observed to arise from regions that cover
at least 3 percent of the disc's surface area.  \citet{Shahbaz03b}
have suggested that the large ($\sim$ few hrs) flares in V404\,Cyg are
produced in regions further out or further above the disc from a corona than
the rapid ($\sim$0.5\,hr) and more rapid ($\sim$\,min) flares.

If the flares in V404\,Cyg and \target\ have the same origin, then  it is
interesting to note that the large flares ($\sim$ few hrs) observed in
V404\,Cyg cover a larger surface area of the disc compared to the rapid 
(tens of mins) flares observed in \target, which is consistent with the idea
that the large flares arise from regions that extend further out into the 
disc compared to the rapid flares.  Although the flare temperatures derived
suggest that large flares are cooler than rapid flares, the uncertainties
are  large and so no meaningful conclusion can be drawn at this stage.
Accurate physical parameters for the flares can only be obtained  by
resolving the Balmer jump and Paschen continuum.

\subsection{Do the flares arise from the bright-spot?}

For the data presented  here on \target,  there is evidence that  some of the
flare events in the continuum  lightcurve are correlated with the Balmer
emission line lightcurves, similar to what is observed in 
V404\,Cyg \citep{Hynes02}.
The value for the Balmer decrement  suggests that the persistent flux is
optically thin and the decrease of the Balmer decrement during the flares
suggests a significant temperature increase.
We find that the optically thin spectrum of the
flare, which lasts tens of minutes, has a temperature of $\sim$12000\,K and
covers 0.08 percent of the disc's projected surface area 
(see section\,\ref{FLARE}).

In many high inclination systems  the continuum light from the bright-spot
produces a single hump in the lightcurve, because the bright-spot is obscured
by the disc when  it is on the side of the disc facing away from the observer. 
Since \target\ is at an intermediate inclination angle
(41$^\circ$; \citealt{Shahbaz94} and \citealt{Gelino01}) where no strong
obscuration effects are predicted, it is not surprising that the continuum
lightcurve does not show obvious evidence of a bright-spot. From the Doppler
maps of \target\ (see section\,\ref{BSPOT}), it is  clear that the bright-spot
is present in the  Balmer emission lines. If the bright spot were present in
the continuum then we would expect strong correlations between the continuum
and emission line lightcurves. Indeed we see some evidence for such 
correlations in the lightcurves (see Figures\,\ref{FIG:VLT_LCURVES} and
\ref{FIG:WHT_LCURVES}). It is
interesting to note that the flare temperature and area we obtain  (see
section\,\ref{FLARE}) is consistent with the bright-spot temperature
and radius determined by \citet{Gelino01}.
Therefore, with the data presented here we cannot rule out the possibility that
the flares arise from the bright-spot. Although, this hypothesis can only be tested by
observing high inclination X-ray  transients, where one would expect the
orbital distribution of the flares and  lightcurves to be indicative, 
it should be noted that the kinematics of the flares in V404\,Cyg do 
rule out at least some of the flares coming from the bright spot.

\section{CONCLUSION}

We have presented a time-resolved spectrophotometric  quiescent study of the
optical variability in \target.  We observe the well known flare events which
and  find that some of the events appear in both the
continuum and  the emission line lightcurves.  The Balmer line flux and
variations suggest that that the persistent emission is optically thin and  the
drop in the Balmer decrement for the flare event at phase 1.15, suggests that
either a significant increase in temperature occurred during the flare or that
the flare is more optically thick than the continuum. 
The data suggests that there are two HI emitting regions, the accretion disc 
and the accretion stream/disc region, with different Balmer decrements. 
The orbital modulation of H$\alpha$ with the continuum suggests that the
steeper decrement is most likely associated with the stream/disc impact region.

We find that flare spectrum can be described by an optically thin gas with a
temperature in the range 10000--14000\,K that covers 0.05--0.08 percent (90
percent confidence) of the accretion disc's surface.  Given these parameters,
the possibility that the flares arise from the bright-spot cannot be ruled out.

Finally, Doppler images of the H$\alpha$ and H$\beta$  emission show enhanced 
emission at the stream/disc impacts  region as well as extended structure from  
the opposite side of the disc.  The trailed spectra show characteristic S-wave 
features that can be interpreted in the context of an  eccentric accretion disc.

\section*{Acknowledgements}
TS acknowledges support from the Spanish Ministry of Science and Technology 
under project AYA\,2002\,03570.
RIH is supported by NASA through Hubble Fellowship grant 
\#HF-01150.01-A\ awarded by the Space Telescope Science Institute, which 
is operated by the Association of Universities for Research in Astronomy, 
Inc., for NASA, under contract NAS 5-26555.
Based on observations made with ESO Telescopes at the Paranal Observatory 
under programme 70.D-0766 and also with  the William Herschel Telescope  
operated on the island of La Palma by the Issac Newton Group
in the Spanish Observatorio del
Roque de los Muchachos of the Instituto de Astrof\'\i{}sica de Canarias.

\end{document}